# Network meta-analysis of rare events using penalized likelihood regression

**Theodoros Evrenoglou[*1], Ian R. White[2], Sivem Afach[3], Dimitris Mavridis[1,4], Anna Chaimani[1,5]**

[1] *Université de Paris, Research Center of Epidemiology and Statistics (CRESS-U1153), INSERM, Paris, France*

[2] *MRC Clinical Trials Unit, University College London, London, UK*

[3] *Université Paris-Est Créteil, UPEC, EpiDermE EA 7379, Créteil, F-94010, France*

[4] *Department of Primary Education, University of Ioannina, Greece*

[5] *Cochrane, France*

*Corresponding Author:

Theodoros Evrenoglou

Université de Paris, Research Center of Epidemiology and Statistics (CRESS-U1153), INSERM, Paris, France

Hôpital Hôtel-Dieu, 1 Place du Parvis Notre-Dame, 75004 Paris

email: tevrenoglou@gmail.com

**Abstract**: Network meta-analysis (NMA) of rare events has attracted little attention in the literature. Until recently, networks of interventions with rare events were analyzed using the inverse-variance NMA approach. However, when events are rare the normal approximation made by this model can be poor and effect estimates are potentially biased. Other methods for the synthesis of such data are the recent extension of the Mantel-Haenszel approach to NMA or the use of the non-central hypergeometric distribution. In this article, we suggest a new common-effect NMA approach that can be applied even in networks of interventions with extremely low or even zero number of events without requiring study exclusion or arbitrary imputations. Our method is based on the implementation of the penalized likelihood function proposed by Firth for bias reduction of the maximum likelihood estimate to the logistic expression of the NMA model. A limitation of our method is that heterogeneity cannot be taken into account as an additive parameter as in most meta-analytical models. However, we account for heterogeneity by incorporating a multiplicative overdispersion term using a two-stage approach. We show through simulations that our method performs consistently well across all tested scenarios and most often results in smaller bias than other available methods. We also illustrate the use of our method through two clinical examples. We conclude that our 'penalized likelihood NMA' approach is promising for the analysis of binary outcomes with rare events especially for networks with very few studies per comparison and very low control group risks.

**Keywords**: rare endpoints, bias reduction, multiple treatment meta-analysis, maximum likelihood estimates

# 1 Introduction

Network meta-analysis (NMA) has become an essential tool of comparative effectiveness research[1]. Despite the rapid and extensive development of NMA methods over the last decade[2], little attention has been given to the issue of rare events within a network of interventions. Typically, NMAs with rare events are performed using the standard 'inverse-variance' (IV) model with a continuity correction for studies with zero events. A major drawback of this approach is that it relies on large sample approximations and normality assumptions which are implausible under the presence of low number of observed events. A popular alternative approach, usually fitted in Bayesian framework, is based on the exact binomial distribution. However, it has been suggested that Bayesian methods may be problematic for meta-analyses of rare events since results can be dominated by the prior distribution[3,4]. On the other hand, when such a model is fitted in frequentist framework it relies on the maximum likelihood estimation which is known to be biased in the presence of rare events[5,6]. Other, possibly less biased, options for performing NMA with rare events are the recent extension of the Mantel-Haenszel (MH) meta-analytical model[4] and a model based on the non-central hypergeometric distribution (NCH)[4]; to date these NMA models are only common-effect and do not allow for heterogeneity of treatment effect between studies.

A further important issue of most meta-analyses with rare events is how to handle studies with zero events in all treatment groups. Imputations of arbitrarily chosen constants (e.g 0.5) have been found to seriously bias the results[7,8]. More sophisticated models (e.g. random intercepts models) allow inclusion of such studies by using between-study information which, however, may bias the results[9,10]. Existing methods for NMA with rare events that avoid the use of between-study information, like the aforementioned MH and NCH models, either exclude these studies from the analysis or require continuity corrections. Studies with zero events in all treatment groups are not infrequent in meta-analyses of rare endpoints and the optimal way to treat such studies is still unclear. Salpenter et al.[11] published a systematic review of 148 studies on the incidence of lactic acidosis with metformin use in type 2 diabetes mellitus; all the studies in the review reported zero events. A recent meta-epidemiological study of 442 Cochrane reviews including at least one study with no events in both treatment groups suggested that the inclusion or exclusion of these studies can impact materially the results of the meta-analyses[12]. In addition, in the context of NMA, exclusion of these studies may lead to disconnected networks.

In the present paper, we aim to tackle the problem of rare events in networks of interventions with binary data by adapting and extending methodology from the analysis of individual studies. We use the logistic regression expression of NMA[13] and the penalization to the likelihood function proposed by Firth[14] to estimate the NMA relative effects, using odds ratios, through a one-stage model. In this way, we attempt to likely obtain less biased and more precise estimates in comparison to the existing methods for NMAs with rare events described earlier. On top of that, our penalized likelihood NMA method (PL-NMA) is the first approach that, without using between-study information, allows the meta-analysts to choose between including or excluding from the analysis studies with zero events in all treatment groups The rest of the article is structured as follows. In section 2, we describe the standard NMA model as a logistic regression model and the proposed PL-NMA model. Section 3 presents a simulation study exploring the performance of our method in terms of bias and precision in comparison to existing NMA approaches for rare events under different scenarios. Finally, in Section 4 we apply the different models in two exemplar NMA datasets, and in Section 5 we discuss the strengths and limitations of our PL-NMA model.

## 2  Methods

### 2.1  NMA as logistic regression model

Suppose a network of $N$ studies and $T$ treatments. Let $r_{ik}$ and $n_{ik}$ denote the number of events and the number of total participants respectively in treatment group $k$ of study $i$ where $i = 1,2, \dots N$ and $k \in K_i$ with $K_i = \{treatments\ evaluated\ in\ study\ i\}$. We assume that

$$r_{ik} \sim Bin(n_{ik}, p_{ik})$$

where $p_{ik}$ is the probability of an event in treatment group $k$ of study $i$. Then, we model the probabilities $p_{ik}$ through the following common-effect logistic regression model,

$$logit(p_{ik}) = \alpha_i + X_{ik}d_{b_ik} \quad (1)$$

with $b_i$ an arbitrarily chosen baseline treatment from the set $K_i$ and

$$X_{ik} = \begin{cases} 1, if\ k \neq b_i \\ 0, if\ k = b_i \end{cases}$$

Hence, for $k \neq b_i$, $d_{b_ik}$ represents the log-odds ratio ($logOR$) of treatment $k$ versus $b_i$ in the $i^{th}$ study. The parameter $a_i$ represents the log-odds of the event in the $b_i$ group and it is treated as a nuisance parameter. Assuming consistency, the $logOR$ between any two treatments $t_1$ and $t_2$ ($t_1, t_2 = 1, ..., T$ and $t_1 \neq t_2$) is obtained as

$$d_{t_1 t_2} = d_{b_i t_2} - d_{b_i t_1}$$

It follows from (1) that,

$$p_{ik} = expit(a_i + X_{ik}d_{b_ik})$$

Let $\boldsymbol{\alpha}$ to denote the vector that contains all study intercepts $\alpha_i$, similarly let $\boldsymbol{d}$ to denote the vector of the relative effects $d_{b_ik}$ and finally let $\boldsymbol{r}$ and $\boldsymbol{n}$ to denote the vectors of observed events and sample sizes per study and treatment arm.

The estimation of $p_{ik}$ relies on the maximization of the likelihood function which can be written as,

$$L(\boldsymbol{\alpha}, \boldsymbol{d}|\boldsymbol{r}, \boldsymbol{n}) = \prod_{i=1}^{N} \prod_{k \in K_i} \binom{n_{ik}}{r_{ik}} expit(\alpha_i + X_{ik}d_{b_ik})^{r_{ik}} \left(1 - expit(\alpha_i + X_{ik}d_{b_ik})\right)^{n_{ik} - r_{ik}}$$

Equivalently, the log-likelihood function is,

$$l(\boldsymbol{\alpha}, \boldsymbol{d}|\boldsymbol{r}, \boldsymbol{n}) = \sum_{i=1}^{N} \sum_{k \in K_i} log\binom{n_{ik}}{r_{ik}} + r_{ik} log(expit(\alpha_i + X_{ik}d_{b_ik})) \quad (2)$$

$$+ (n_{ik} - r_{ik}) log(1 - expit(\alpha_i + X_{ik}d_{b_ik}))$$

The model in equation (1) can be extended to the so-called Binomial-Normal (BN-NMA) model if we replace the $d_{b_ik}$ with $\delta_{i,b_ik}$ which are now the study specific treatment effects that are assumed to follow a normal distribution with mean $d_{b_ik}$ and variance $\tau^2$. In this model, the true treatment effects vary from study to study but the study intercepts $\alpha_i$ are kept fixed. The model can also allow for random intercept terms $a_i$ if we further assume that these also follow a normal distribution.

## 2.2 Common-effect penalized likelihood NMA

To reduce the bias of the above MLE for NMAs with rare events, we employ Firth's modification[14] to the log-likelihood function in (2); this is the frequentist equivalent to using as penalty Jeffrey's invariant prior. This modification results in the penalized likelihood and log-likelihood functions respectively,

$$L^*(\boldsymbol{\alpha}, \boldsymbol{d}|\boldsymbol{r}, \boldsymbol{n}) = L(\boldsymbol{\alpha}, \boldsymbol{d}|\boldsymbol{r}, \boldsymbol{n}) \; |\mathbf{I}|^{\frac{1}{2}}$$

and

$$l^*(\boldsymbol{\alpha}, \boldsymbol{d}|\boldsymbol{r}, \boldsymbol{n}) = l(\boldsymbol{\alpha}, \boldsymbol{d}|\boldsymbol{r}, \boldsymbol{n}) + \frac{1}{2} \log |\mathbf{I}| \qquad (3)$$

where the matrix $\mathbf{I} \equiv \mathbf{I}(\boldsymbol{\alpha}, \boldsymbol{d})$ is the Fisher's information matrix that is given as

$$\mathbf{I} = (\mathbf{Z}'\mathbf{W}\mathbf{Z})^{-1} \qquad (4)$$

with $\mathbf{Z}$ being the model's design matrix with dimensions $\sum_{i=1}^{N} A_i \times (N + T - 1)$ and entries 1 for the columns associated with the relevant treatment and study and 0 otherwise[16]. $A_i$ is the number of arms in study $i$, $\mathbf{W} = \boldsymbol{diag}\{n_{ik}p_{ik}(1-p_{ik})\}$ and $|\mathbf{I}|$ is the determinant of $\mathbf{I}$. After maximizing the penalized likelihood function, we can replace $p_{ik}$ with $\hat{p}_{ik}$ in matrix $\mathbf{W}$. Then, the square roots of the diagonal elements of the $\mathbf{I}$ matrix in (4) represent the standard errors of the estimates which can be used to construct Wald type confidence intervals. An alternative option is to construct profile likelihood confidence intervals that can be obtained using the critical region defined by the following inequality,

$$2\{l^*(\hat{\boldsymbol{\alpha}}, \hat{\boldsymbol{d}}|\boldsymbol{r}, \boldsymbol{n}) - l^*(\hat{\boldsymbol{\alpha}}, \hat{\boldsymbol{d}}_0|\boldsymbol{r}, \boldsymbol{n})\} \leq c_{1,1-q}$$

where $\hat{\boldsymbol{d}}_0 = (d_{b_ik_0}, \hat{\boldsymbol{d}}_{b_ik})$ is the vector that contains the estimated treatment effects of all treatments in the network versus the reference treatment, except $k_0 \neq k$ which is a specific treatment in the network; finally $c_{1,1-q}$ is the $(1-q)$ quantile of the chi-square distribution with 1 degree of freedom.

In case of $T = 2$, the model in (1) with the likelihood function in (3) corresponds to the penalized likelihood regression model for pairwise meta-analysis.

The above PL-NMA provides finite treatment effects and standard errors even in the case of studies that report zero events in all treatment arms. Specifically, the issue of studies reporting only zero events in meta-analysis resembles that of separation (i.e. when one or more covariates perfectly predict the outcome) in individual studies[17,18]. The PL-NMA method allows to gain information from studies with zero events in two or more groups without sharing information between the studies.

## 2.3  Random-effects penalized likelihood NMA

The above PL-NMA model is a common-effect model and its extension to incorporate heterogeneity is challenging. In particular, the standard approach of incorporating heterogeneity as an additive term is not straightforward as the penalization requires closed forms of the likelihood function and its moments and these are not available for a logistic regression mixed-effect model. Therefore, we use an alternative approach previously suggested in the literature where heterogeneity is incorporated as a multiplicative term[19,20]. This is a two-stage approach that modifies the study variances through an overdispersion parameter. Specifically, after fitting the model presented in the previous section, we multiply the study variances with a scale parameter $\varphi$[21-23]. Hence,

$$v_{ik}^* = v_{ik}\varphi, \varphi \geq 1$$

where $v_{ik}$ are the within-study variances. In this way, the point estimates of the treatment effects remain unaffected but their variances are inflated by that unknown parameter $\varphi$[21]. To estimate the parameter $\varphi$ we use the expression suggested by Fletcher[24]

$$\hat{\varphi} = \frac{\hat{\varphi}_P}{1 + \bar{s}}$$

where $\bar{s}$ is the mean value of $s_{ik} = \left(\frac{\partial \hat{v}_{ik}}{\partial \hat{p}_{ik}}\right)\frac{(r_{ik}-\hat{E}(r_{ik}))}{\hat{v}_{ik}}$, $\frac{\partial \hat{v}_{ik}}{\partial \hat{p}_{ik}}$ is the derivative of $\hat{v}_{ik}$ with respect to $\hat{p}_{ik}$ and $\hat{\varphi}_P = \frac{P}{m}$ with $P$ denoting the Pearson's statistic and $m = (T-1) + (N-1)$ the residual degrees of freedom of the model. The Pearson's statistic for the NMA model presented in the previous section is,

$$P = \sum_{i=1}^{N}\sum_{k \in K_i} \frac{r_{ik} - \hat{E}(r_{ik})}{\hat{v}_{ik}}$$

If $\hat{\varphi} < 1$, we set it to 1.

## 3 Simulations

We conducted a simulation study to compare our proposed and existing methodologies for meta-analysis of rare events. We considered a total of 32 scenarios with the first 16 scenarios assuming the presence of only two-arm studies. We report our simulation following the recommendations by Morris et al.[25]

### 3.1 Data generation

We start the data generation process by specifying the number of treatments in the network (3, 5 or 8) and the number of studies in every treatment comparison (2, 4 or 8). We always create all possible $\frac{T(T-1)}{2}$ treatment comparisons; hence we only assume fully-connected networks. We then generate the total number of participants per study arm using a uniform distribution, namely $n_{ik} \sim Unif(c_1, c_2)$, with $c_1 = 30, c_2 = 60$ for generating small studies and $c_1 = 100, c_2 = 200$ for larger studies. We only generated studies with an equal number of participants across arms. For each study we generated the control group risk; that is the risk of the event for patients receiving the reference treatment irrespective of whether this was evaluated in the study, thus for the simulations $b_i \equiv 1$. We generate the risk of an event in the reference treatment group 1 using $p_{i1} \sim Unif(u_1, u_2)$, with $u_1 \in [0.5\% - 5\%], u_2 \in [1\% - 10\%]$ depending on the scenario

(see **Table 1**). In all scenarios, the true $logOR$s were fixed at equal intervals between 0 and 1 (e.g. in a network of 5 treatments the true $logOR$s are set to 0.25, 0.5, 0.75 and 1 for the comparisons of treatments 2, 3, 4 and 5 versus the reference treatment 1, respectively) and were used to calculate the odds of an event in every study arm,

$$odds_{i1} = \frac{p_{i1}}{1 - p_{i1}}$$

$$odds_{ik} = odds_{i1} * OR, k \neq 1$$

Finally, we obtain the event probabilities in the non-reference treatment groups as $p_{ik} = \frac{odds_{ik}}{1+odds_{ik}}$ and the number of events in every study arm using a binomial distribution $r_{ik} \sim Bin(n_{ik}, p_{ik}), \forall i, k$.

For scenarios assuming the presence of heterogeneity, we set a common parameter $\tau$ to represent the standard deviation of the random-effects (see the Supplementary material for a description).

We explored a total of 32 scenarios with varying baseline risks of the event in the studies, number of treatments in the network, study size, number of studies per comparison, and magnitude of heterogeneity (**Table 1**). In scenarios 1-16 we considered only two-arm studies while in scenarios 17-32 we only included multi-arm studies evaluating all treatments. The latter are certainly not very realistic scenarios but they aimed at exploring whether the increased correlation from multi-arm studies could contribute to precision and bias reduction in NMAs with rare events. Scenarios 1-10 only consider small studies (i.e. 30-60 participants per arm), with varying small baseline risks. We considered both homogeneous and heterogeneous cases and with different number of studies per comparison. For scenarios 11-32, we increased the number of patients per arm and we mostly focus on cases in which the event risks were extremely low (i.e. 1-2% and 0.5-1%). For the scenarios involving multi-arm studies (17-32), we generated studies with 100 to 200 participants per arm and both homogeneous and heterogeneous samples. Finally, in Scenarios 21-24 and 29-32 we considered a mix of extremely low with higher baseline risks. **Table 1** provides a summary of the different scenarios.

**Table 1:** Overview of scenarios examined in our simulation. For each scenario we generated 1000 datasets. The rows in bold represent the scenarios with multi-arm studies.

| # | Treatments in the network | Patients per arm | Number of studies per comparison | Total number of studies per dataset | Heterogeneity (τ) | Control group risk (%) | Mean events per study |
|---|---|---|---|---|---|---|---|
| 1 | 5 | 30-60 | 2 | 20 | 0 | 3-5% | 3 |
| 2 | 5 | 30-60 | 2 | 20 | 0.1 | 3-5% | 3 |
| 3 | 5 | 30-60 | 2 | 20 | 0 | 5-10% | 6 |
| 4 | 5 | 30-60 | 2 | 20 | 0.1 | 5-10% | 6 |
| 5 | 5 | 30-60 | 4 | 40 | 0 | 3-5% | 3 |
| 6 | 5 | 30-60 | 4 | 40 | 0.1 | 3-5% | 3 |
| 7 | 5 | 30-60 | 4 | 40 | 0 | 5-10% | 6 |
| 8 | 5 | 30-60 | 4 | 40 | 0.1 | 5-10% | 6 |
| 9 | 8 | 30-60 | 2 | 56 | 0 | 3-5% | 3 |
| 10 | 8 | 30-60 | 2 | 56 | 0.1 | 3-5% | 3 |
| 11 | 5 | 100-200 | 2 | 20 | 0 | 1-2% | 4 |
| 12 | 5 | 100-200 | 2 | 20 | 0.1 | 1-2% | 4 |
| 13 | 5 | 100-200 | 2 | 20 | 0 | 0.5-1% | 2 |
| 14 | 5 | 100-200 | 2 | 20 | 0.1 | 0.5-1% | 2 |
| 15 | 5 | 100-200 | 4 | 40 | 0 | 0.5-1% | 2 |
| 16 | 5 | 100-200 | 4 | 40 | 0.1 | 0.5-1% | 2 |
| **17** | **3** | **100-200** | **8** | **8** | **0** | **1-2%** | **4** |
| **18** | **3** | **100-200** | **8** | **8** | **0.1** | **1-2%** | **4** |
| **19** | **3** | **100-200** | **8** | **8** | **0** | **0.5-1%** | **2** |
| **20** | **3** | **100-200** | **8** | **8** | **0.1** | **0.5-1%** | **2** |
| **21** | **3** | **100-200** | **8** | **8** | **0** | **0.5-5%** | **7** |
| **22** | **3** | **100-200** | **8** | **8** | **0.1** | **0.5-5%** | **7** |
| **23** | **3** | **100-200** | **8** | **8** | **0** | **0.5-10%** | **13** |
| **24** | **3** | **100-200** | **8** | **8** | **0.1** | **0.5-10%** | **13** |
| **25** | **5** | **100-200** | **8** | **8** | **0** | **1-2%** | **4** |
| **26** | **5** | **100-200** | **8** | **8** | **0.1** | **1-2%** | **4** |
| **27** | **5** | **100-200** | **8** | **8** | **0** | **0.5-1%** | **2** |
| **28** | **5** | **100-200** | **8** | **8** | **0.1** | **0.5-1%** | **2** |
| **29** | **5** | **100-200** | **8** | **8** | **0** | **0.5-5%** | **7** |
| **30** | **5** | **100-200** | **8** | **8** | **0.1** | **0.5-5%** | **7** |
| **31** | **5** | **100-200** | **8** | **8** | **0** | **0.5-10%** | **13** |
| **32** | **5** | **100-200** | **8** | **8** | **0.1** | **0.5-10%** | **13** |

Our simulation did not investigate the impact of studies with zero events in all treatment arms. **Table 2** shows that across all scenarios the majority of the 1000 simulated datasets included only a handful of studies reporting only zero events. The latter implies that any impact of such studies on the final results cannot be captured through the present simulation study.

**Table 2:** Extent of studies with zero events in all treatment arms in the simulation study.

| Scenario | Minimum | 1st Quartile | Median | 3rd Quartile | Maximum | Total number of studies per dataset |
|---|---|---|---|---|---|---|
| 1 | 0 | 0 | 0 | 0 | 3 | 20 |
| 2 | 0 | 0 | 0 | 0 | 2 | 20 |
| 3 | 0 | 0 | 0 | 0 | 1 | 20 |
| 4 | 0 | 0 | 0 | 0 | 1 | 20 |
| 5 | 0 | 0 | 0 | 0 | 3 | 40 |
| 6 | 0 | 0 | 0 | 1 | 4 | 40 |
| 7 | 0 | 0 | 0 | 0 | 1 | 40 |
| 8 | 0 | 0 | 0 | 0 | 1 | 40 |
| 9 | 0 | 0 | 0 | 1 | 3 | 56 |
| 10 | 0 | 0 | 0 | 1 | 3 | 56 |
| 11 | 0 | 0 | 0 | 0 | 2 | 20 |
| 12 | 0 | 0 | 0 | 0 | 1 | 20 |
| 13 | 0 | 0 | 1 | 1 | 4 | 20 |
| 14 | 0 | 0 | 1 | 1 | 5 | 20 |
| 15 | 0 | 1 | 1 | 2 | 6 | 40 |
| 16 | 0 | 1 | 1 | 2 | 6 | 40 |
| 17 | 0 | 0 | 0 | 0 | 1 | 8 |
| 18 | 0 | 0 | 0 | 0 | 0 | 8 |
| 19 | 0 | 0 | 0 | 0 | 1 | 8 |
| 20 | 0 | 0 | 0 | 0 | 2 | 8 |
| 21 | 0 | 0 | 0 | 0 | 1 | 8 |
| 22 | 0 | 0 | 0 | 0 | 1 | 8 |
| 23 | 0 | 0 | 0 | 0 | 1 | 8 |
| 24 | 0 | 0 | 0 | 0 | 1 | 8 |
| 25 | 0 | 0 | 0 | 0 | 0 | 8 |
| 26 | 0 | 0 | 0 | 0 | 0 | 8 |
| 27 | 0 | 0 | 0 | 0 | 2 | 8 |
| 28 | 0 | 0 | 0 | 0 | 1 | 8 |
| 29 | 0 | 0 | 0 | 0 | 1 | 8 |
| 30 | 0 | 0 | 0 | 0 | 1 | 8 |
| 31 | 0 | 0 | 0 | 0 | 0 | 8 |
| 32 | 0 | 0 | 0 | 0 | 1 | 8 |

## 3.2 Evaluated models and estimands

In each scenario, we generated 1000 datasets and we compared the following models:

- Common- and random-effects IV-NMA model with 0.5 correction for studies with at least one zero event arm
- Common-effect MH-NMA and NCH-NMA without 0.5 correction for studies with zero event arms.
- Common- and random-effects PL-NMA model.
- Common effect logistic regression NMA model with standard (unpenalized) likelihood.
- Random-effects logistic regression NMA model using MLE with both fixed and random intercept(s);
- Common- and random-effects Bayesian NMA with exact binomial likelihood and non-informative priors to the treatment effects, $d_{b_ik} \sim N(0, 100^2)$, and heterogeneity, $\tau^2 \sim Unif(0,4)^{26}$. The random-effects model is the Bayesian equivalent of the frequentist BN-NMA model. We run 2 chains with 50.000 iterations and discarded the first 10.000 samples from each chain.

A brief description of all models is provided in the Supplementary material.

## 3.3 Estimands and performance measures

The estimands of interest are the $T - 1$ $logOR$s between treatments $t = 2, ..., T$ and treatment 1. The performance of the models was investigated in terms of the mean bias defined as the mean difference between the estimated and the true $logOR$s averaged over simulated datasets and estimands for each NMA method: the true estimands are all positive so averaging over estimands makes sense. We further calculated the mean coverage in each scenario defined as the percent of the corresponding 95% confidence (or credible) intervals that included the corresponding true $logOR$. Finally, the methods were compared in terms of the mean squared error (MSE) and mean length of confidence (or credible) intervals averaged across multiple contrasts and simulated datasets. The simulations were conducted using R version 4.0.3 (2020-10-10) and the packages netmeta[27], brglm[28,29], lme[28], and gemtc[31]. The code used for our simulations can be found here https://github.com/TEvrenoglou/PLNMA-simulation.

## 3.4 Simulation results

In **Figure 1**, we present the results in terms of mean bias across multiple contrasts (multiple estimands). In most scenarios, IV-common effect and IV-random-effects gave the most biased results. The two models usually yielded very similar results and estimated heterogeneity as 0 no matter if the true value of heterogeneity parameter $\tau$ was set to 0 or 0.1. The two PL-NMA models overall produced the least biased results in most scenarios. The MH-NMA, NCH-NMA, and BN-NMA methods also performed generally well, but the MH-NMA and NCH-NMA resulted in large bias in scenarios with many treatments in the network and a small number of studies per comparison (i.e. scenarios 13, 14). The performance of NCH model for the scenarios with higher risks (scenario 3, 4, 7, 8) was decreased as this model makes use of assumptions that are valid only for very low risks of the event (see the Supplementary material for details). Interestingly, the PL-NMA models had a stable performance across the different scenarios in terms of bias with a maximum value of mean bias equal to 0.02. The BN-NMA models provided satisfactory results in terms of bias in almost all scenarios with a maximum mean bias of 0.06. The performance of BN-NMA with fixed intercept was increased in scenarios involving relatively large studies, namely when participants per arm were between 100 and 200. The results of the common-effect logistic regression model with the standard (unpenalized) likelihood were equivalent to the results of the BN-NMA model with fixed intercept. The BN-NMA model with random intercepts had a better performance in terms of bias in comparison to the BN-NMA model with fixed intercept. The model provides unbiased results in many cases and when some amount of bias exists this is in general very small. This model had some convergence issues but failures were very rare across the 1000 datasets of each scenario. An increase in convergence failures was observed in the scenarios with 8 treatments in the network (Supplementary Table 1) The Bayesian common-effect model had better performance than the respective random-effects model, but there are many scenarios where results from both models are suffering from an important amount of bias.

In all scenarios, the results were not affected materially for any of the models when the data-mechanism involved heterogeneity. Full results are available in the Supplementary Tables 2 and 3 and are consistent with the results that we obtained for each contrast separately (Supplementary Figures 1-4).

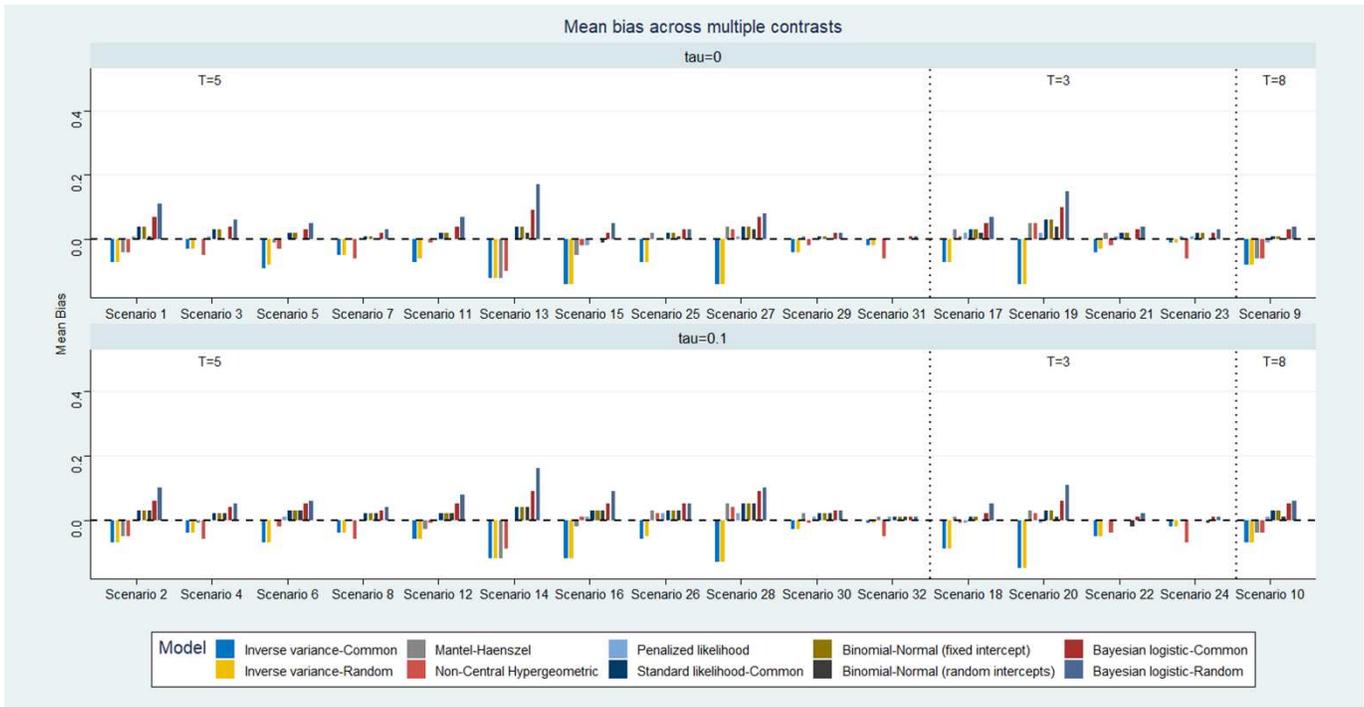

**Figure 1:** Simulation results in terms of mean bias for scenarios withwith 5, 3 and 8 treatments (T) respectively. Missing bars correspond to 0 mean bias (after rounding to two decimal places) for the respective NMA method. The Monte-Carlo standard error across the different scenarios and methods ranges from 0.004 to 0.02 with a mean value equal to 0.01.

In terms of estimation of the heterogeneity, the Bayesian random-effects model appeared to have the worst performance; in all scenarios it substantially overestimated $\tau$ (heterogeneity results available only in the Supplementary Table 4). Of course, this can be explained by the fact that the prior distribution is centered at 2 while the true value is 0.1 or 0. The IV-random–effects model seemed to have a good performance for estimating $\tau$ apart from scenario 16 where 3 treatments many studies per comparison were available; in that case, the model estimated that $\tau = 0$ across all 1000 simulations. The BN model with fixed intercept had a poor performance in estimating $\tau$ as in all heterogeneous scenarios the heterogeneity parameter was estimated being 0. This explains why the results of this model were equal to those obtained from the corresponding common-effect model with the standard likelihood. The BN-model with random intercepts appeared to have a better capacity to estimate $\tau \neq 0$, for heterogeneous scenarios. The mean bias for estimating $\tau$ in those scenarios ranged from -0.04 to -0.09. Finally, since there is no clear correspondence between $\tau$ and $\varphi$ we calculated for $\varphi$ the number of cases for which $\varphi > 1$ across all datasets and all heterogeneous scenarios. For scenarios with higher control group risks (scenarios 4,8, 22, 24, 30, 32) we observed that $\varphi$ was greater than one in 20% to 40% of the simulated datasets. However, this frequency was reduced as the event risks were becoming lower and 0 for scenario 16.

In terms of coverage probability, the PL-NMA models did not perform very well when the Wald-type confidence intervals were used (**Figure 2**). The coverage probability was increased, though, by using profile likelihood confidence intervals and in most scenarios, it was above the nominal level of 95%. MH-NMA and NCH-NMA models had similarly good performance in terms of coverage probability close to the nominal level, but it should be noted that the performance of NCH model was again decreased for the scenarios with higher event risks (scenarios 3, 4, 7, 8). Overall, for IV-NMA, MH-NMA, and NCH-NMA results in terms of coverage are consistent with previous simulation studies[4]. The common-effect Bayesian NMA model appeared to have the same issues as the (frequentist) BN-NMA with fixed intercept whose performance was problematic in terms of coverage. The BN-NMA model with random intercepts was found to suffer from important under-coverage in scenarios with very low risks (e.g. 13, 14, 15, 16, 19, and 20), the performance of the method was increased for scenarios with higher risks. On the other hand, the random-effects Bayesian model had in almost all scenarios the best performance across all the other models in terms of coverage probability.

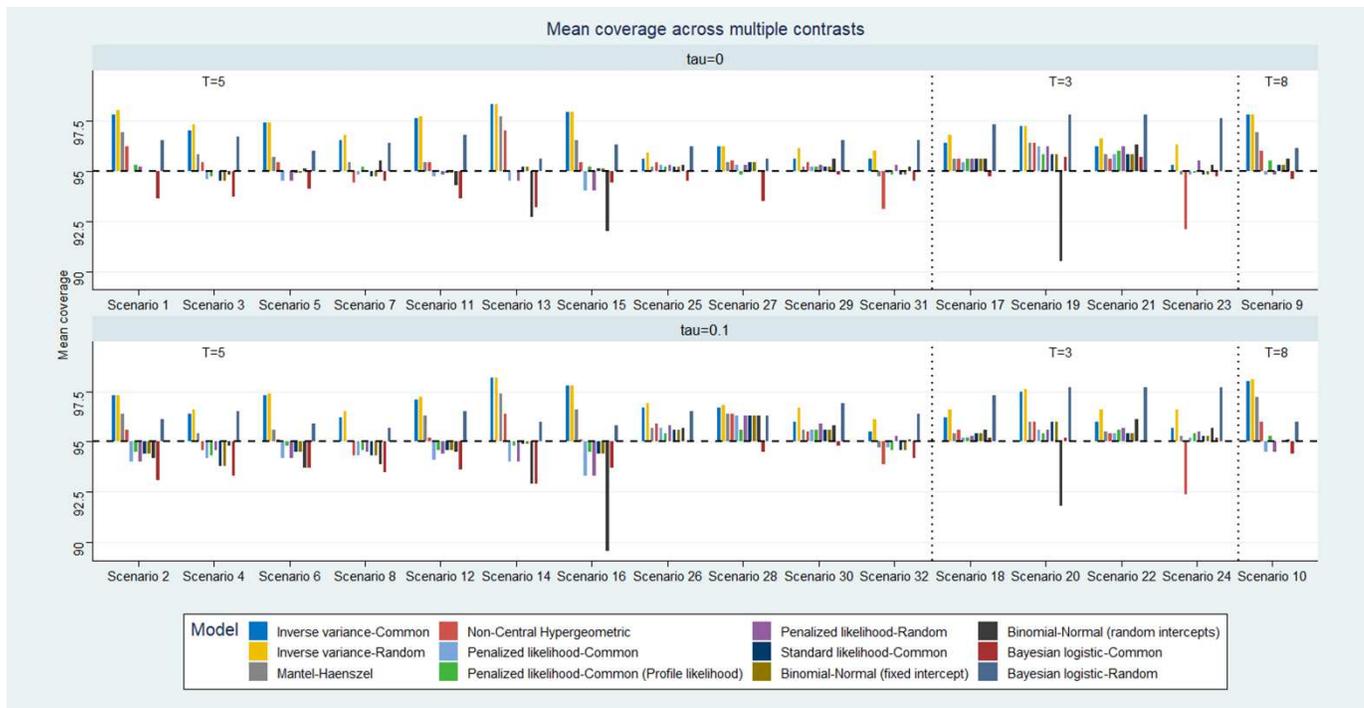

**Figure 2:** Simulation results in terms of coverage probability for scenarios with with 5, 3 and 8 treatments (T) respectively.

In terms of MSE, the BN-NMA model with random intercepts provided slightly better results in comparison to the other models (**Figure 3**). An exception is the random-effects Bayesian model which gave the larger MSE with some extreme cases (scenarios 13, 14) where the average MSE was twice the average MSEs reported by the other models. Finally, regarding the mean length of confidence (or credible intervals) the Wald-type confidence intervals obtained by the BN-NMA with random intercepts and the PL-NMA method were found to have the smaller lengths in comparison to the other models (**Figure 4**). The mean length of the profile-likelihood confidence intervals for PL-NMA, though, were wider; this probably explains also their better performance compared to the Wald-type intervals in terms of coverage probability. The most uncertain results across the 32 were obtained by the Bayesian random-effects model which provided the widest credible intervals.

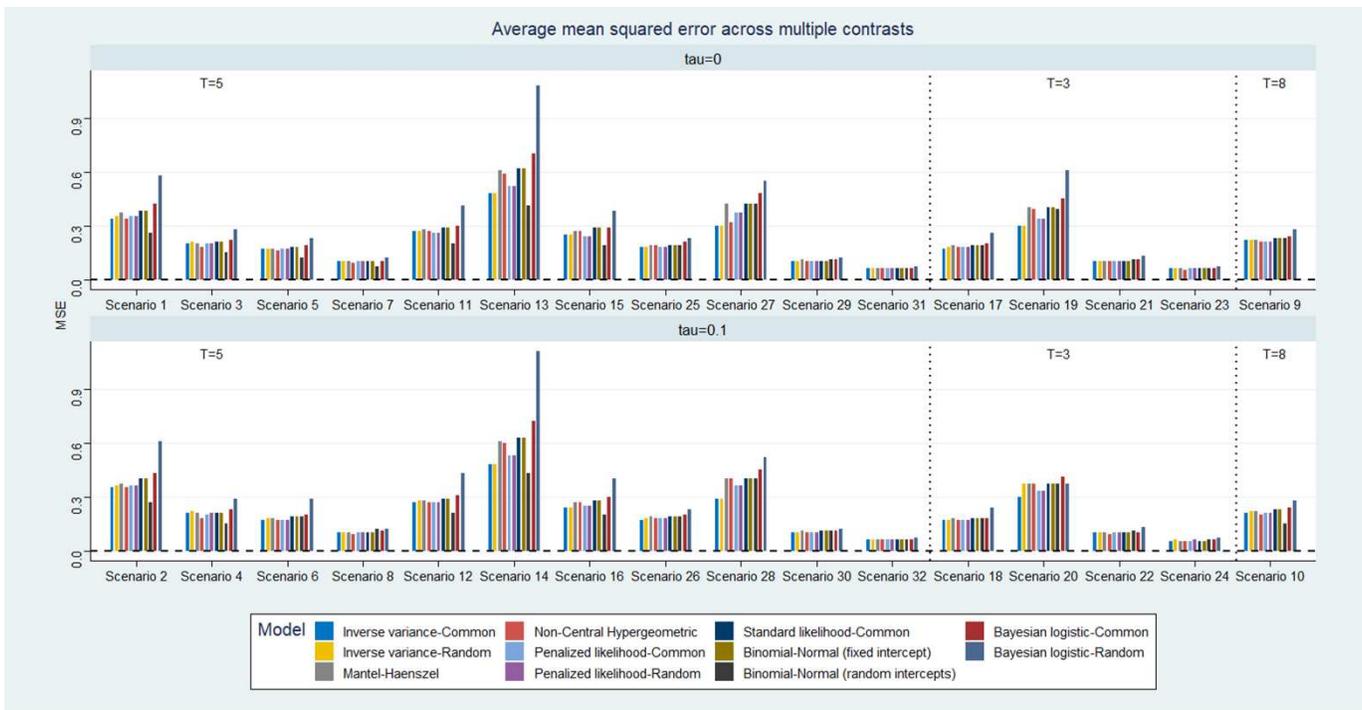

**Figure 3:** Simulation results in terms of mean squared error (MSE) for scenarios with with 5, 3 and 8 treatments (T) respectively.

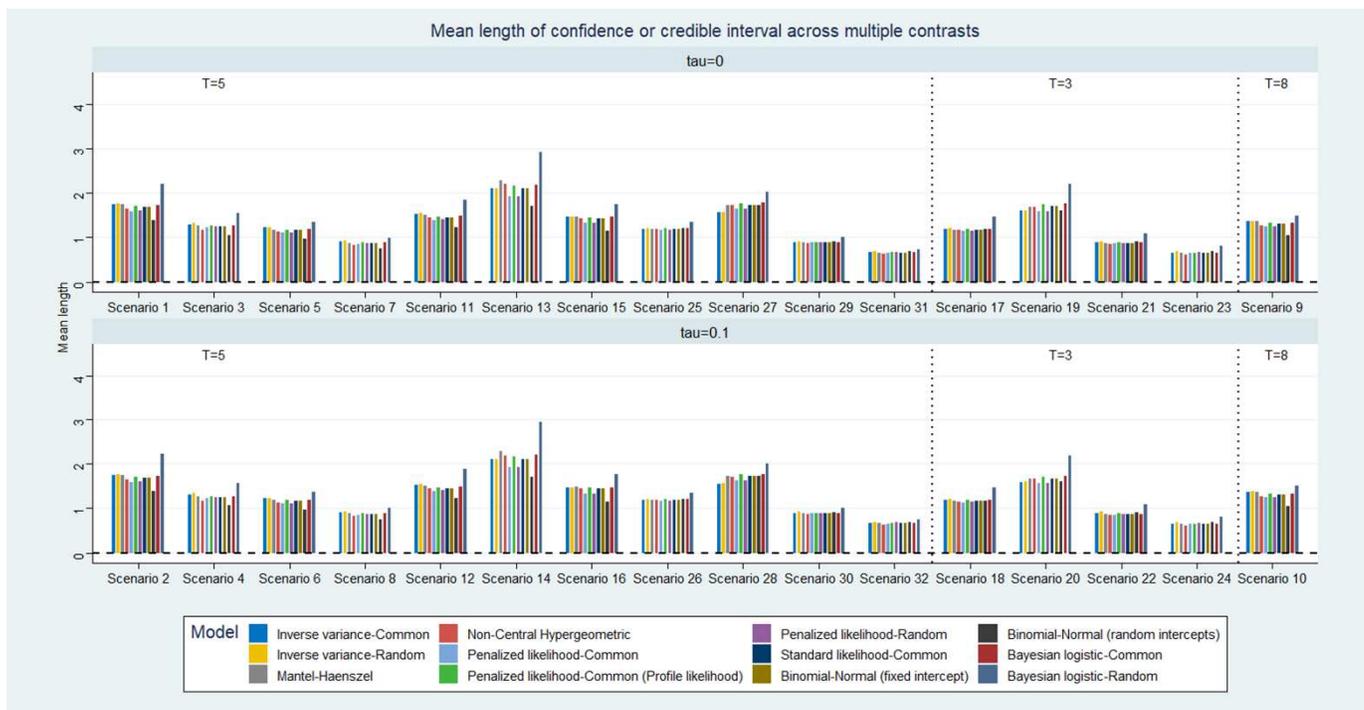

**Figure 4:** Simulation results in terms of length of confidence (or credible) intervals for scenarios with with 5, 3 and 8 treatments (T) respectively.

## 4 Clinical examples

### 4.1 Safety of inhaled medications for patients with chronic obstructive pulmonary disease

We compared the results across the different models using a network that evaluates the safety of inhaled medications for patients with chronic obstructive pulmonary disease[32]. The network consists of 41 studies and the outcome of interest is mortality. This is supposed to be a rare outcome as from 52462 patients in total only 2408 (4.6%) experienced the event of interest. In total, 13 out of 41 studies reported less than 5 events and 20 out of 99 arms had 0 events. The network diagram of this dataset is depicted in Erreur ! Source du renvoi introuvable.. We analyzed the data using the PL-NMA model both with Wald and profile likelihood CIs, the MH-NMA model, the NCH-NMA model, the BN-NMA model, and finally both the common and random-effects Bayesian models. For the latter convergence was checked using the Brooks-Gelman-Rubin[33] criterion. We did not use the IV model due to the bad performance of the method in the simulations.

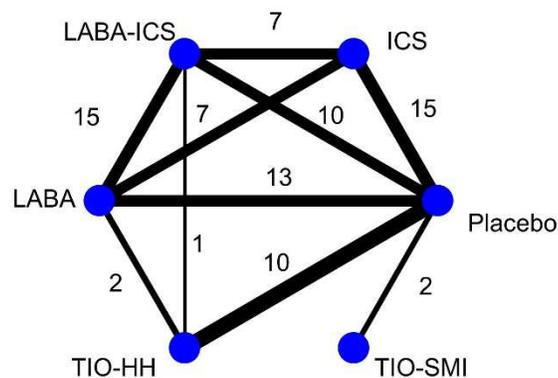

**Figure 5:** Network diagram for the inhaled medications example. (ICS=inhaled corticosteroid, LABA=long-acting β2 agonist, TIO-HH=tiotropium dry powder).

The results across the different NMA methods are almost identical (**Figure 6**). Regarding the estimation of heterogeneity, the additive parameter $\tau$ was estimated as zero for the BN-NMA model and the multiplicative parameter $\varphi$ was estimated as 1. However, the Bayesian random-effects model resulted in $\tau= 0.12$ [0.005, 0.36].

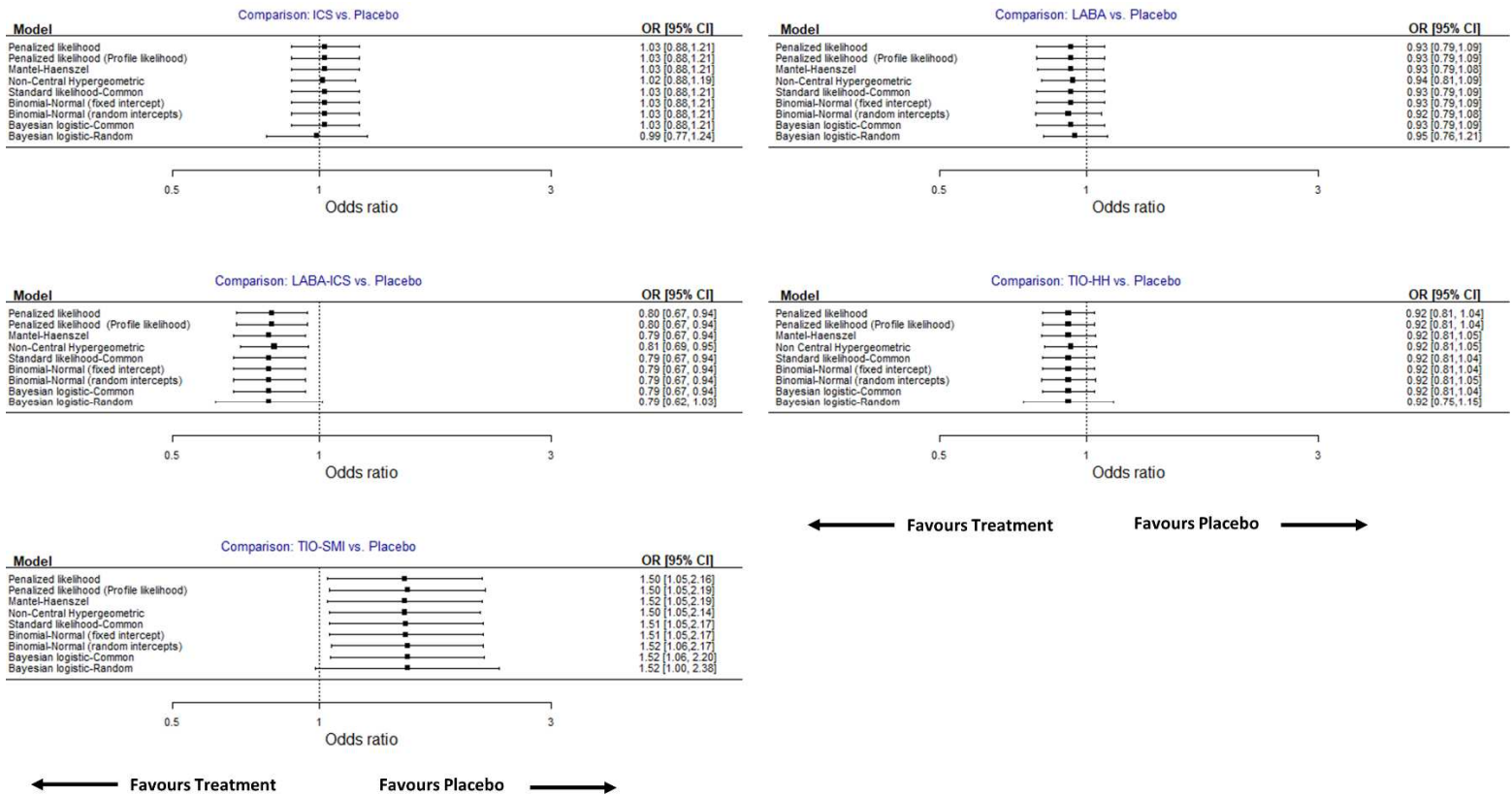

**Figure 6:** Forest plots showing the odds ratios obtained from the inhaled medications network for all comparisons against the reference.

## 4.2 Safety of different drug classes for chronic plaque psoriasis

The second example is a network that evaluates the safety of different interventions for chronic plaque psoriasis[34]. The dataset consists of 43 studies that involve 5 drug classes and placebo. Here, we compared again the different NMA methods but in a more extreme situation where the event risks in the studies range between 0 and 1%. The outcome of interest is the number of malignancies that occurred after using the drugs. The mean sample size per study arm is 226 patients. Out of the 43 studies in this network, 29 (69%) are studies with zero events in all treatment groups. All other methods than the PL-NMA and BN-NMA with random intercepts cannot take into account these studies. For the MH-NMA and NCH-NMA methods, the initially connected network becomes disconnected since the comparisons related to the class Anti-IL 23 needs to be removed (**Figure 7a** and **Figure 7b**).

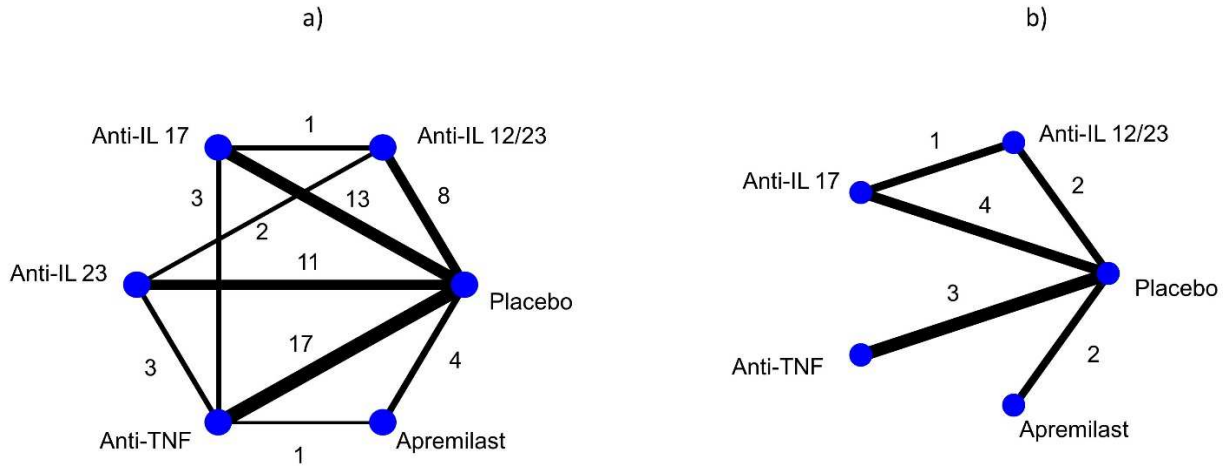

**Figure 7:** Network diagrams for the psoriasis example. Panel (a) shows the initially well-connected network while panel (b) the resulting disconnected network after the exclusion of studies that report only zero events. (Anti-TNF=Anti-tumor necrosis factor, Anti-IL=Anti-interleukin).

The results of the different approaches are shown in **Figure 8**. For all comparisons, the PL-NMA model that included all available studies in the network (i.e. 43 studies) gave the most precise relative effect estimates. Excluding the studies with zero events in all arms slightly reduced the precision of the PL-NMA but it remained higher than the precision of the other methods. The BN-NMA model with random intercepts was found to be more sensitive in the choice of inclusion or exclusions. The differences in the results are observed both in terms of the point estimation as also in the estimation of the uncertainty. In terms of point estimates, for two out of the five comparisons in **Figure 8**, all models gave very similar results and for one comparison (Anti-TNF vs Placebo) the MH-NMA and NCH-NMA gave somewhat different estimates than the other approaches. With respect to heterogeneity, the BN-NMA with fixed intercept model estimated the additive term $\tau$ as 0 and the PL-NMA the corresponding multiplicative term $\varphi$ as 1. So, both models suggested the absence of statistical heterogeneity. The BN-NMA model with random intercepts gave estimated $\tau$ as 0.14, but when the studies with all zero event arms were excluded the heterogeneity parameter was estimated as 0. The random-effects Bayesian model, though, gave $\hat{\tau}$ = 0.97 (0.05,1.94).

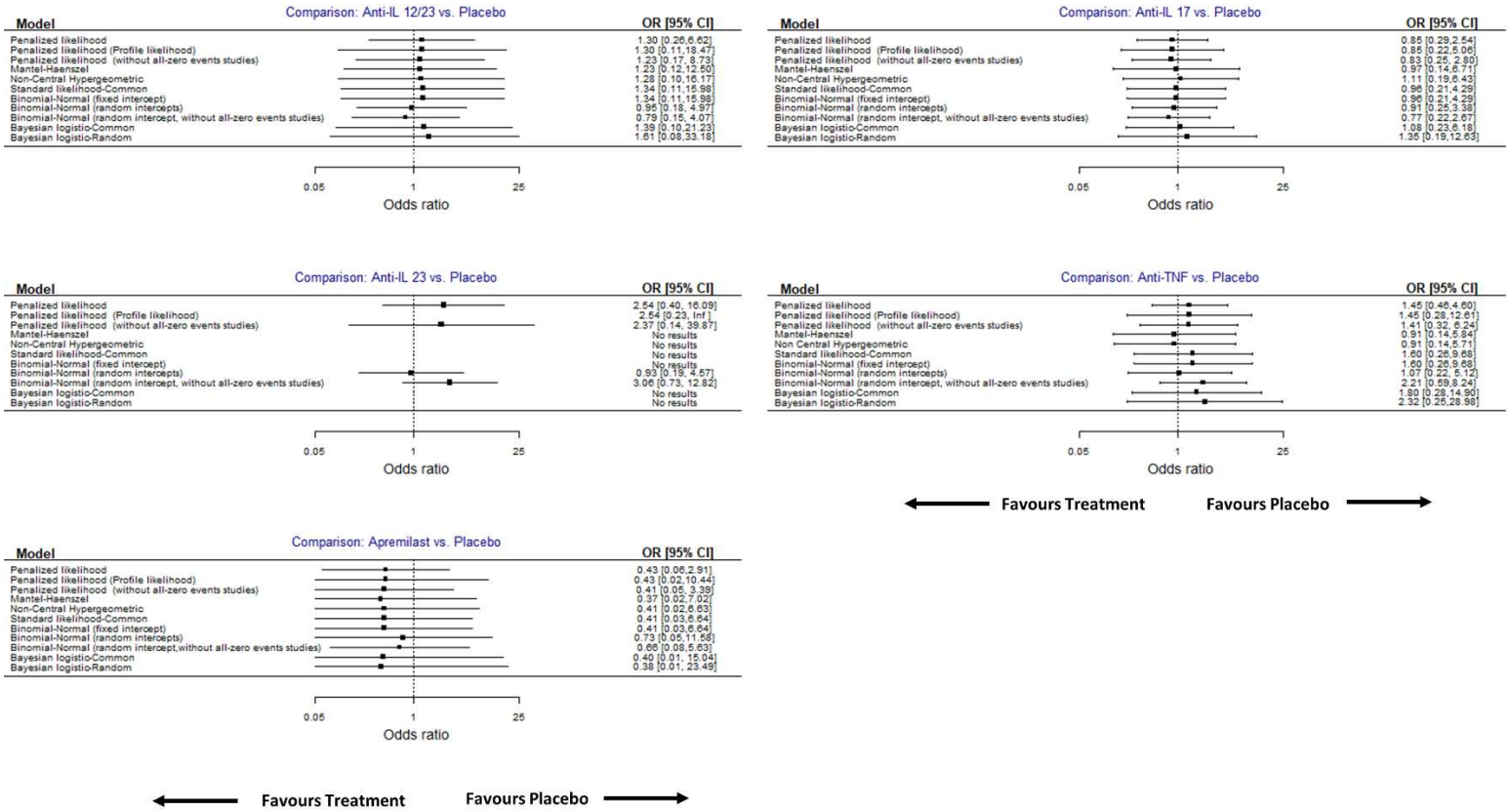

**Figure 8:** Forest plots showing the odds ratios obtained from psoriasis network for all comparisons against the reference.

## 5 Discussion

In this paper, we have presented a new method for NMA of binary outcomes and rare events. Our method aims to improve the performance of existing NMA approaches for rare events in terms of bias and precision by using the penalized likelihood function for logistic regression that was originally proposed by Firth[14] for individual studies. Our approach can only provide odds ratios. However, in the context of rare events, the differences between odds ratios and risk ratios are often negligible[35,36].

We evaluated the performance of the different methods and compared their results through an extended simulation study and two real clinical examples. We used various scenarios including studies with low or extremely low event risks. Our proposed PL-NMA model appeared to have overall the best performance in terms of bias especially in situations with very few studies per comparison and very low control group risks. The method, though, performed somewhat poorly in terms of coverage probability with Wald-type confidence intervals. The performance on coverage probability was substantially improved when profile-likelihood confidence intervals were used but precision was reduced. In agreement with previous simulation studies[4], the IV-NMA models are considered a suboptimal choice and we believe meta-analysts should avoid their use for NMAs with rare events, especially for the cases with very small number of events per study (e.g. <3) . MH-NMA, NCH-NMA are good options under certain conditions, but they become less reliable as the network becomes sparser. BN-NMA models are shown to be more robust in terms of bias. Finally, the common-effect Bayesian performed well in terms of bias but was poor in terms of coverage whereas the random-effects Bayesian model had good coverage but poor performance in terms of bias. As expected, all models failed to sufficiently frequently detect the presence of heterogeneity particularly for scenarios with extremely low event risks (i.e. 0.5-1% and 1-2%).

An additional property of the proposed PL-NMA model is the ability to synthesize all studies within a network of interventions irrespective of the number of events per arm since excluding studies from the analysis may result in disconnected networks. The problem lies mainly in networks including studies with zero events in two or more treatment groups. To date, the optimal way to treat such studies in meta-analysis or NMA is still unclear with some researchers arguing that they are informative[12,37] and others that those studies are non-informative and problematic[4,36]. Our method provides the flexibility of allowing both inclusion or exclusion of these studies. In the psoriasis example, which involved several such studies, we observed that they may improve the precision of the estimated odds ratios. Other methods for including zero-event studies do so by incorporating between-study information; further work is needed to explore whether this is also true for the PL-NMA approach.

Although the PL-NMA model is by nature a common-effect model, we used a two-stage approach for incorporating between-study variance as a multiplicative overdispersion parameter. Multiplicative parameters are not the standard way to account for heterogeneity in meta-analysis as typically an additive parameter is implemented. However, the penalized likelihood approach strongly relies on the analytical expression of the likelihood function and its moments; these are not available for the logistic regression model of Section 2.1 in the case of random-effects. Thus, the incorporation of an additive heterogeneity parameter in the PL-NMA method is not straightforward. The use of multiplicative heterogeneity parameters, though, has been suggested previously for some meta-analytical approaches[19,20] and particularly for meta-analysis of rare events by Kuss[37] and by Kulinskaya and Olkin[23]. The latter, in contrast to our approach, is a one-stage random-effects model[23]. A key difference between our two-stage approach and the usual one-stage random effects models is that the former does not affect the relative effect estimates but only inflates their variance when $\varphi > 1$. In addition, the incorporation of the multiplicative parameter does not affect the weights of the studies and hence results might be dominated by the larger studies[22,38]. In the simulations, we found that all random-effects approaches did not perform sufficiently well with respect to the estimation of the heterogeneity. It should be noted, though, that the Cochrane handbook suggests that estimation of heterogeneity is not of primary interest when performing meta-analyses of rare events and the priority should be the estimation of the treatment effects[36].

Overall, the proposed PL-NMA model offers a reliable choice for performing NMA for binary outcomes with rare events. Future work for NMA of rare events could consider the extension of the beta-binomial model for meta-analysis into NMA[37]. Meta-analysts should always bear in mind that the presence of studies with rare events makes estimation challenging and, therefore, a sensitivity analysis should be carried out to investigate the robustness of the results under various analysis schemes. It is planned to integrate the proposed PL-NMA method into the R package netmeta.

**Acknowledgements:** The authors would like to thank Dr Gerta Rücker for providing useful comments on an earlier version of the manuscript. Ian White was supported by the Medical Research Council Programme MC_UU_00004/06.

**Data Availability Statement:** Data and R codes for the analysis of the two clinical examples can be found in the Supplementary material.